%% file: usp4-final.tex
\documentclass[12pt, epsfig]{article}                            
\usepackage{epsfig}       
\usepackage{amssymb}     
\usepackage{amsfonts}    
\newskip\humongous \humongous=0pt plus 1000pt minus 1000pt

\newif\ifdtup

\catcode`\@=11
 
\@addtoreset{equation}{section}
\def\theequation{\thesection.\arabic{equation}}
 
\def\@normalsize{\@setsize\normalsize{15pt}\xiipt\@xiipt
\abovedisplayskip 14pt plus3pt minus3pt%
\belowdisplayskip \abovedisplayskip
\abovedisplayshortskip \z@ plus3pt%
\belowdisplayshortskip 7pt plus3.5pt minus0pt}
 
\def\small{\@setsize\small{13.6pt}\xipt\@xipt
\abovedisplayskip 13pt plus3pt minus3pt%
\belowdisplayskip \abovedisplayskip
\abovedisplayshortskip \z@ plus3pt%
\belowdisplayshortskip 7pt plus3.5pt minus0pt
\def\@listi{\parsep 4.5pt plus 2pt minus 1pt
     \itemsep \parsep
     \topsep 9pt plus 3pt minus 3pt}}
 
\relax

\catcode`@=12
 
\catcode`\@=11
 
\def\section{\@startsection{section}{1}{\z@}{3.5ex plus 1ex minus
   .2ex}{2.3ex plus .2ex}{\large\bf}}
 
\def\thesection{\arabic{section}}    
\def\thesubsection{\arabic{section}.\arabic{subsection}}

\def\appendix{\setcounter{section}{0}
 \def\thesection{Appendix \Alph{section}}
 \def\thesubsection{\Alph{section}.\arabic{subsection}}
 \def\theequation{\Alph{section}.\arabic{equation}}}
 
\def\YGrule{0.4}   
\def\YGbox{6.5}    
\def\SymBoxes#1#2#3#4{\newdimen\un@t \un@t#3%
\raisebox{#1}{\rule{#2\un@t}{#4}\hskip-#2\un@t
\@tempdimb\un@t \advance\@tempdimb by-#4\@tempcntb#2\relax%
\@whilenum{\@tempcntb>0}\do{
\rule{#4}{\un@t}\hskip\@tempdimb \advance\@tempcntb by\m@ne}%
\hskip-#2\un@t \rule[\un@t]{#2\un@t}{#4}%
\rule[\un@t]{#4}{#4}\hskip-#4
\rule{#4}{\un@t}}\hskip-#4}                
\def\Young{\@ifnextchar[{\@Young}{\@Young[0]}}
\def\@Young[#1]#2{\newdimen\YG@unit \YG@unit\YGbox pt%
\newdimen\h@ight \h@ight#1\YG@unit \@tempcnta-1\relax
\@tfor\c@ount:=#2\do{\advance\@tempcnta by\@ne}
\@tempdima\@tempcnta\YG@unit%
\advance\h@ight by\@tempdima\relax     
\@tfor\c@ount:=#2\do{\SymBoxes{\h@ight}{\c@ount}{\YG@unit}{\YGrule pt}%
\@tempdima-\c@ount\YG@unit \hskip\@tempdima%
\advance \h@ight by -\YG@unit}         
\@tempdima\YG@unit \multiply\@tempdima by\@car#2\@nil %
\hskip\@tempdima}                      
\def\YoungTab{\@ifnextchar[{\@YoungIdx}{\@YoungIdx[0]}}
\def\@YoungIdx[#1]{\@ifnextchar[{\@iYoungIdx[#1]}{\@iYoungIdx[#1][\@empty]}}
\def\@iYoungIdx[#1][#2]#3{%
\newdimen\YG@unit \YG@unit\YGbox pt\newdimen\YG@rule \YG@rule \YGrule pt
\newcount\c@ount \c@ount\z@ \newdimen\skip@wd \unitlength\@ne pt
\newdimen\h@ight \h@ight#1\YG@unit \@tempcnta\m@ne\relax
\@tfor\d@um:=#3\do{\advance\@tempcnta by\@ne}
\@tempdima\@tempcnta\YG@unit%
\advance\h@ight by\@tempdima\relax
\@tfor\@idxlist:=#3\do{
\@tempcnta\z@\hskip.5\YG@rule\relax 
\@for\@idx:=\@idxlist\do{
\raisebox{\h@ight}{\makebox(\YGbox,\YGbox){#2$\@idx$}}
\advance\@tempcnta by\@ne}\hskip-.5\YG@rule%
\@tempdima-\@tempcnta\YG@unit \hskip\@tempdima%
\ifnum\c@ount=\z@ \skip@wd-\@tempdima\fi \relax
\SymBoxes{\h@ight}{\@tempcnta}{\YG@unit}{\YG@rule}%
\hskip\@tempdima \advance\h@ight by -\YG@unit
\advance\c@ount by\@ne}
\hskip\skip@wd}                      

\begin{document}

\newcommand{\beq}{\begin{equation}}
\newcommand{\eeq}{\end{equation}}
\newcommand{\bea}{\begin{eqnarray}}
\newcommand{\eea}{\end{eqnarray}}
\newcommand{\beas}{\begin{eqnarray*}}
\newcommand{\eeas}{\end{eqnarray*}}
\newcommand{\defi}{\stackrel{\rm def}{=}}
\newcommand{\non}{\nonumber}
\newcommand{\bquo}{\begin{quote}}
\newcommand{\enqu}{\end{quote}}
\def\de{\partial}
\def\Tr{ \hbox{\rm Tr}}
\def\const{\hbox {\rm const.}}
\def\o{\over}
\def\im{\hbox{\rm Im}}
\def\re{\hbox{\rm Re}}
\def\bra{\langle}\def\ket{\rangle}
\def\Arg{\hbox {\rm Arg}}
\def\Re{\hbox {\rm Re}}
\def\Im{\hbox {\rm Im}}
\def\diag{\hbox{\rm diag}}
\def\longvert{{\rule[-2mm]{0.1mm}{7mm}}\,}
\begin{titlepage}
{\hfill     IFUP-TH/2004-04} 
\bigskip
\bigskip

\begin{center}
{\large  {\bf
SUPERCONFORMAL VACUA \\ IN $N=2$ $USp(4)$ GAUGE THEORIES
 } }
\end{center}

\bigskip
\begin{center}
{\large  Roberto AUZZI $^{(2,3)}$ and Roberto GRENA $^{(1,3)}$
 \vskip 0.10cm
 }
\end{center}

\begin{center}
{\it   \footnotesize

Dipartimento di Fisica ``E. Fermi" -- Universit\`a di Pisa $^{(1)}$, \\
Scuola Normale Superiore - Pisa $^{(2)}$,
 Piazza dei Cavalieri 7, Pisa, Italy \\
Istituto Nazionale di Fisica Nucleare -- Sezione di Pisa $^{(3)}$, \\
     Via Buonarroti, 2, Ed. C, 56127 Pisa,  Italy $^{(1,3)}$ \\
 }
{\bf  auzzi@sns.it, grena@df.unipi.it \\}

\end {center}

\noindent  
{\bf Abstract:}

 { We study the dynamics of a confining vacuum in $N=2$ $USp(4)$
   gauge theory with $n_f=4$. 
   The vacuum appears
   to be a deformed conformal theory with nonabelian
   gauge symmetry. The low-energy degrees of freedom
   consist of four nonabelian magnetic monopole doublets of the
   effective $SU(2)$ colour group, two dyon doublets and one electric doublet.
   In this description the flavour quantum number is carried
   only by the monopoles.
   We argue that confinement is caused by the condensation
   of these monopoles, and involves strongly interacting
   nonabelian degrees of freedom.
  }

\vfill  
 
\begin{flushright}
   February 2004
\end{flushright}
\end{titlepage}

\bigskip

\hfill{}
\bigskip

\section{Introduction}

$N=2$ gauge theories have been a consistent source of hints as to the nature of real-world QCD. Different types of confining vacua are realized
in these models. For example some models exhibit confinement due to the condensation
of monopoles charged under the maximally abelian subgroup, as in the
$N=1$ vacua surviving the adjoint mass perturbation in $N=2$ SYM
\cite{SW1,SW2}.

However, this is not the typical situation in softly broken $N=2$ theories
with fundamental matter fields.
In \cite{APS}, \cite{CKM}, \cite{CKKM} the mechanism of confinement and
dynamical symmetry breaking has been studied in some detail
in these theories.
In some vacua the low-energy degrees of freedom turn out to be
{\it nonabelian monopoles} of the type studied by
Goddard-Nuyts-Olive \cite{GNO}; these objects
(see \cite{BK}, \cite{monopoli} for a semiclassical analysis)
also carry a flavor quantum number and their condensation
is responsible for confinement and dynamical flavor symmetry breaking.

The most interesting type of vacua found in \cite{CKM} is based
on the deformation of interacting superconformal field theories (\cite{AD},\cite{SCFT},\cite{Eguchi}).
The low-energy dynamics involves relatively non-local monopoles and dyons, as
in the case first discovered by Argyres and Douglas \cite{AD}.
In a previous paper \cite{AGK} we have studied  one of these confining vacua
in $N=2$ $SU(3)$ gauge theories with $N_f=4$ quark hypermultiplets.
Upon a small adjoint mass perturbation confinement occurs,
as can be demonstrated
through various considerations, such as the study of
 the large adjoint mass limit (see \cite{CKM}).
The low energy degrees of freedom consist of four
$SU(2)$ gauge doublets monopoles, one dyon doublet and one electric doublet.
These relatively nonlocal particles conspire
to give a vanishing beta function, generalizing the abelian Argyres-Douglas mechanism
\cite{AD} to a nonabelian effective theory.
Giving the quark
hypermultiplets different masses, the superconformal nonabelian vacuum
split into six abelian local vacua;
the fact that monopole condensates vanished in the equal mass limit suggested
that confinement occurred  in an essentially different way, involving
strongly interacting nonabelian magnetic monopoles.

Confinement is described in this way in many
vacua in $N=2$ supersymmetric gauge theories. Given the importance of the problem and subtlety of the nonabelian confinement mechanism, we believe that it is worthwhile to analyze as many explicit examples as possible, to deepen our insight into this phenomena further.
In this paper we pursue the analysis for a vacuum
in the $USp(4)$ theory with $N_f=4$.
The degrees of freedom are found to be the same as in \cite{AGK}, plus another dyon.
This is a little puzzling because the
 $\beta$-function cancellation mechanism does not seem to work in the same
 way as in the previously studied $SU(3)$ case.

Finally, we make a conjecture as to which degrees
of freedom condense, by using the pattern of symmetry breaking known from the large adjoint mass analysis (see \cite{CKM}); we conclude
again that nonabelian monopoles play a fundamental role both in confinement and in dynamical symmetry breaking.

 \section{$SU(2)$  Vacua in a $USp(4)$, $n_f=4$ Gauge Theory}

The Seiberg-Witten curve of the $USp(4)$ theory (\cite{curves}), with $n_f$ quarks of different masses $m_a$, is (setting $\Lambda=1$)
\begin{equation}
xy^2=[x(x^2-Ux-V)+2\prod_{a=1}^{n_f}m_a]^2-4\prod_{a=1}^{n_f}(x+m_a^2),
\end{equation}
where
\begin{equation}
U=\phi_1^2+\phi_2^2,\hspace{20pt}V=-\phi_1^2\phi_2^2.
\end{equation}
$\phi_1$ and $\phi_2$ are the two components of the adjoint scalar field that breaks the gauge symmetry:
\begin{equation}
\langle \phi \rangle=\left(\begin{array}{cccc}\phi_1&0&0&0\\0&\phi_2&0&0\\0&0&-\phi_1&0\\0&0&0&-\phi_2\end{array}\right).
\end{equation}
In the case $m_a=0$, the behavior of the curve is highly singular at the two points $U=\pm 2$, $V=0$,
 where four of the five branchpoints coalesce. In these vacua we have $\phi_1=0$, and so the gauge
symmetry is broken to $USp(2) \times U(1) \simeq SU(2) \times U(1)$.
  Giving equal masses to all of the quarks ($m_a=m$),
  these two singularities exhibit different behaviors: the $U=2$ singularity splits into two separate singularities,
   each of which has two pairs of branchpoints colliding ($U(1)^2$ unbroken).
    The $U=-2$ singularity splits into three singularities: two of them have an unbroken $U(1)^2$,
     while the third has four colliding branchpoints (\cite{CKM}).

Giving distinct masses to the quarks, each of two singularities near $U=2$ splits into four,
 and the $SU(2)$ singularity near $U=-2$ splits into six; the former is a $4+4^*$ flavor representation
 of $U(4)$, the residual global symmetry, the latter is a $1+6+1$ representation.
 The $U=\pm2$, $V=0$ points are octet vacua of the original $SO(8)$ flavor symmetry.

We will analyze the $m_a=0$ case, and try to understand the structure of the low-energy
degrees of freedom and the mechanism of confinement.

\section{Singularity structure in the massless case}

In $m=0$ case, the curve becomes
\begin{equation}
y^2=x(x^2-Ux-V)^2-4x^3.
\end{equation}
The discriminant of the curve is
\begin{equation}
\Delta_s\Delta_+\Delta_-,
\end{equation}
where
\begin{eqnarray}
&\Delta_s=V^6;\\
&\Delta_+=(4-4U+U^2+4V);\\
&\Delta_-=(4+4U+U^2+4V).
\end{eqnarray}
When two of these factors vanish, the singularity is maximal. $\Delta_s=\Delta_{\pm}=0$ corresponds to the two superconformal points we are considering ($U=\pm 2$, $V=0$).

\begin{figure}[h]
\begin{center}
\epsfig{file=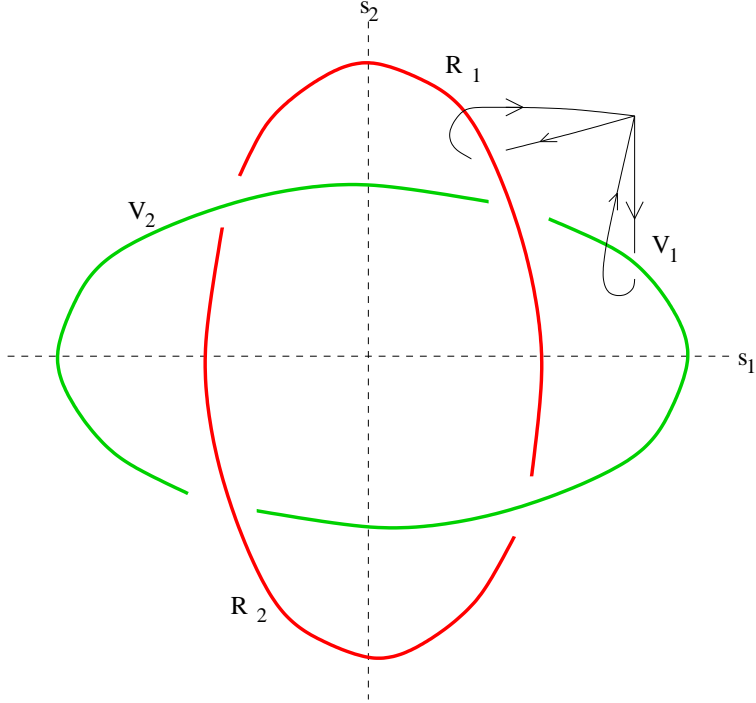, width=10 cm}
\end{center}
\caption{\footnotesize Zero loci of the discriminant of the curve of the
${\cal N}=2,$ $\,USp(4)$, $\,n_f=4$ theory at $m_a=0$.}
\label{tworings}
\end{figure}

Let's consider one of these two points, $U=-2$. It is useful to redefine
\begin{equation}
u=U+2,\hspace{20pt}v=V,
\end{equation}
in order to have $u=0$, $v=0$ in the considered vacuum. At this point, $\Delta_s=\Delta_-=0$. Singularities near this point are located at
\begin{eqnarray}
&\Delta_s=0\longrightarrow v=0;\nonumber\\
&\Delta_-=0\longrightarrow v=-{\displaystyle\frac{u^2}{4}}.
\end{eqnarray}
On the 3-sphere $|u|^2+|v|^2=R^2$ these equations describe two rings with linking number two (fig. \ref{tworings}).

Imposing the vanishing of the first factor ($v=0$), the resulting curve has 3
 branch points colliding; the analysis of monodromies is difficult, because it
  is possible to get charges of the zero-mass particles only when there are pairs
  of branch points colliding ($U(1)^n$ symmetry); this will be clear in the following.


\begin{figure}[h]
\begin{center}
\epsfig{file=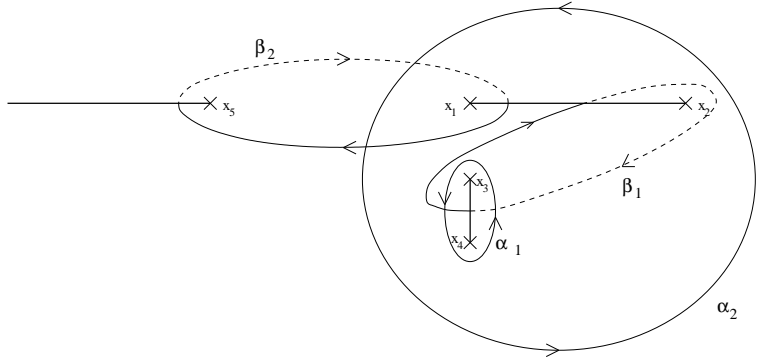, width=10 cm}
\end{center}
\caption{\footnotesize Fundamental cycles considered in the calculation of monodromies.
 $x_1$, $x_2$, $x_3$ and $x_4$ coalesce in the singularity.}
\label{cicli}
\end{figure}

Problems become evident if one considers the monodromies around these two rings,
 as shown in figure \ref{tworings}. Choosing fundamental $\alpha$  and $\beta$
 cycles as in figure \ref{cicli}, one can calculate the non-trivial monodromies
 of these around the singularity rings.  The monodromy $M$ acts on the column vector of
 cycles \beq v=\left(\begin{array}{c} \alpha_1 \\ \alpha_2 \\ \beta_1 \\ \beta_2 \end{array}
 \right)  \eeq
 as $v \rightarrow M.v$. The result is the following:
\begin{eqnarray}
\label{monodm0}
R_1^{m=0}=\left(\begin{array}{cccc}-1&0&0&0\\0&1&0&0\\0&2&-1&0\\-2&0&0&1\end{array}\right), & V_1^{m=0}=\left(\begin{array}{cccc}2&0&1&0\\0&1&0&0\\-1&0&0&0\\0&0&0&1\end{array}\right)\nonumber\\
R_2^{m=0}=\left(\begin{array}{cccc}-1&2&0&0\\0&1&0&0\\0&0&-1&0\\0&0&2&1\end{array}\right), & V_2^{m=0}=\left(\begin{array}{cccc}2&-2&1&0\\0&1&0&0\\-1&2&0&0\\2&-4&2&1\end{array}\right).
\end{eqnarray}
One can easily extract magnetic and electric charges $(g_1,g_2;q_1,q_2)$ from the $V_1, V_2$ monodromies using (\cite{curves}):
\begin{equation}
\label{cariche}
M=\left(\begin{array}{cccc}1+q_1g_1&q_1g_2&q_1^2&q_1q_2\\q_2g_1&1+q_2g_2&q_1q_2&q_2^2\\-g_1^2&-g_1g_2&1-q_1g_1&-q_2g_1\\-g_1g_2&-g_2^2&-q_1g_2&1-q_2g_2\end{array}\right),
\end{equation}
and one obtains the following:
\begin{equation}
V_1=(1,0;1,0),\hspace{20pt}V_2=(1,-2;1,0).
\end{equation}
This is not the case for the $R_1,R_2$ monodromies: they are not in the form (\ref{cariche}).
This is not surprising: at $v=0$ three branch points coalesce, and there is
not a $U(1)^2$ theory: one cannot expect that charges $g_i$, $q_i$ make sense.
 Moreover, it is not a single singularity. The composition of monodromies, when
  two or more singularities coalesce, respects (\ref{cariche}) only if the particles
  are mutually local (for example, if they are in a flavor multiplet). In this case, we
  expect a flavor quartet (magnetic monopoles), plus other two particles; we will see that
  they are not relatively local with respect to the quartet.

\section{Equal mass case and the massless limit}
In order to understand what kinds of particles are present in the theory, we must give a little mass to quarks, in order to split the $v=0$ ring into many simple rings.

We already know that by giving a common mass $m_a=m$ to the quarks, the $U=-2$ singularity splits into three singularities. Making the stereographic projection, one must choose a spherical section of a radius sufficiently large to contain all three singularities: so he can go to the limit $m\rightarrow 0$ without crossing the sphere and altering the topological structure of the singularity rings. Monodromies obtained in such a way remain valid in the limiting case.

The curve becomes
\begin{equation}
xy^2=(x(x^2-(u-2)x-v)+2m^4)^2-4(x+m^2)^4,
\end{equation}
and the discriminant has the three factors:
\begin{eqnarray}
&\Delta_s=(m^4+m^2(u-4)-v)^4,\\
\vspace{5pt}\nonumber\\
&\Delta_1=16m^2+u^2+4v,\\
\vspace{5pt}\nonumber\\
&\Delta_2=432m^8+32m^6(9u-28)+8m^2v(16-8u+u^2-6v)-\nonumber\\
&v^2(u^2-8u+4(v+4))-8m^2(2u^3-22u^2-12(8+v)+u(9v+80)).
\end{eqnarray}
In the singularity, all three of these factors are null. In the $m=0$ case $\Delta_2=v^2\Delta_+$; this gives the $v^6$ factor.

\begin{figure}[h]
\begin{center}
\epsfig{file=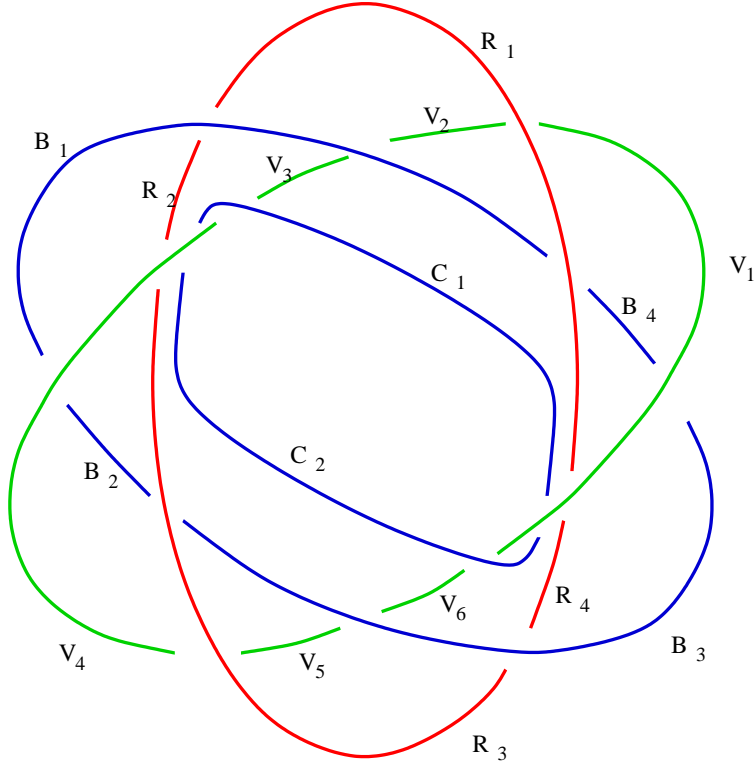, width=10 cm}
\end{center}
\caption{\footnotesize Zero loci of the discriminant of the curve of the
${\cal N}=2,$ $\,USp(4)$, $\,n_f=4$ theory at $m_a=m$, on a $S^3$ sphere containing all the three singularities near $U=-2$, $V=0$.}
\label{fourrings}
\end{figure}

The singularity structure on a 3-sphere of radius $R=1$ has been analyzed numerically, with $m=0.1$. The result are shown in figure \ref{fourrings}: the $R$ ring is a $\Delta_s$ singularity, the $V$ ring is a $\Delta_1$ singularity, the $B$ and $C$ rings are $\Delta_2$ singularities.

First of all, we note that the rings B and C can be continuously deformed to make them coincide with the R ring: this is what happens when $m\rightarrow 0$, and monodromies are not changed by this deformation. This means that this analysis gives correct information on the $m\rightarrow 0$ limit.

We computed non-trivial monodromies around the rings (see \ref{append}). One can explicitly calculate only one monodromy for each ring, and obtain the others from these using the relations shown in \ref{append} (\ref{check}): the redundance of these relations allows a non-trivial check, that works correctly. The charges obtained, with a global sign ambiguity, are:
\begin{eqnarray}
\label{charges}
&R_1=(1,0;0,0)^4;\hspace{20pt}R_2=(0,1;-1,0)^4;\nonumber\\
&R_3=(-1,1;0,0)^4;\hspace{20pt}R_4=(0,0;1,0)^4;\nonumber\\
&B_1=(1,1;-1,0);\hspace{20pt}B_2=(-3,3;-1,0);\nonumber\\
&B_3=(-1,1;1,0);\hspace{20pt}B_4=(3,-1;1,0);\nonumber\\
&C_1=(1,-1;1,0);\hspace{20pt}C_2=(-1,1;-1,0);\nonumber\\
&V_1=V_6=(1,0;1,0);\nonumber\\
&V_4=V_3=(-1,2;-1,0);\nonumber\\
&V_2=(3,0;-1,0);\hspace{20pt}V_5=(-3,2;1,0).
\end{eqnarray}
As we expected, $q_2=0$ for all the charges, because $\beta_2$ cycle remains
 large approaching superconformal point. We expect also that particles relative
  to opposite arches of a singularity circle (ex. $R_1$ and $R_3$) form doublets
   of the unbroken $SU(2)$.
    So one can try to construct a transformation of cycles that give $(g_1,q_1)$
    as the charges of $U(1)\subset SU(2)$, and $(g_2,q_2)$ as the charges of $U(1)$
    orthogonal to $SU(2)$; in this base, $(g_1,q_1)$ of the particles in the doublet
    must have opposite signs, and $(g_2,q_2)$ must have the same sign. It is possible
    to get this applying to charges the transformation
\begin{equation}
T=\left(\begin{array}{cccc}1&0&1&0\\1&2&1&0\\0&0&1&-\frac{1}{2}\\0&0&0&\frac{1}{2}\end{array}\right).
\label{cambio} \end{equation}
Applying $T$, one obtains a $SU(2)\times U(1)$ basis in which the charges are correctly paired into doublets. They are
\begin{eqnarray}
&(R_1,R_3):\hspace{10pt}(\pm1,1;0,0)^4;\nonumber\\
&(R_2,R_4):\hspace{10pt}(\pm1,1;\pm1,0)^4;\nonumber\\
&(B_1,B_3):\hspace{10pt}(0,2;\pm1,0);\nonumber\\
&(B_2,B_4):\hspace{10pt}(\pm4,2;\pm1,0);\nonumber\\
&(C_1,C_2):\hspace{10pt}(\pm2,0;\pm1,0);\nonumber\\
&(V_1,V_4),\,(V_3,V_6):\hspace{10pt}(\pm2,2;\pm1,0);\nonumber\\
&(V_2,V_5):\hspace{10pt}(\pm2,2;\mp1,0).
\end{eqnarray}

\begin{figure}[h]
\begin{center}
\epsfig{file=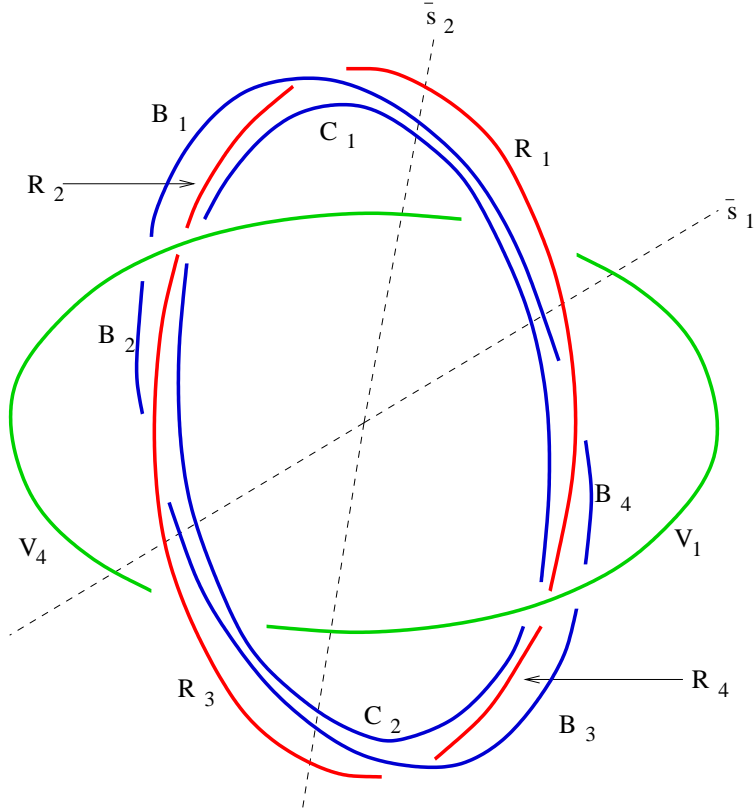, width=10 cm}
\end{center}
\caption{\footnotesize Zero loci of the discriminant of the curve of the
${\cal N}=2,$ $\,USp(4)$, $\,n_f=4$ theory, in the limit $m\rightarrow 0$. The four sections are the same as in fig. \ref{tworings}. }
\label{fourringslimit}
\end{figure}

Following \cite{AGK}, we assume that if we want a correct interpretation of
the theory only a section of fig. \ref{fourrings} must be chosen; other
sections will give equivalent theories.
 In the $m\rightarrow 0$ limit the $R$, $B$, $C$ rings coalesce.
 Observing figure \ref{tworings}, one can see that there are
 only two not equivalent sections ($s_1$ and $s_2$) in this limit.
 The $s_1$ and $s_2$ sections of figure \ref{tworings} can be identified
 with the $\overline{s}_1$ and $\overline{s}_2$ of figure \ref{fourringslimit}.
 The composition of surviving monodromies
  for the rings $R$, $B$, $C$ have to compose into those of (\ref{monodm0}). One can easily check that
\begin{eqnarray}
R_1^{m=0}=C_1R_1B_1;\nonumber\\
R_2^{m=0}=C_2R_3B_3,
\end{eqnarray}
and this shows that $C_1$, $R_1$, $B_1$ are the monodromies that
 compose into $R_1^{m=0}$ of (\ref{monodm0}) in the limit $m\rightarrow 0$; the other monodromies
($R_2$, $R_4$, $B_2$ , $B_4$ , $V_2$, $V_3$, $V_5$ and $V_6$)
 have no role in the massless limit. Thus the sections to consider are $\overline{s}_1$ $\overline{s}_2$
of figure \ref{fourringslimit}.
Corresponding charges relative to $U(1)\subset SU(2)$ are shown in table \ref{bbbb}.
\begin{table}[h]
\begin{center}
\begin{tabular}{|l|l|l|l|}
\hline $\overline{s}_1$ & $\overline{s}_2$ \\
\hline $ R_1,R_3:\,(\pm 1,0)^4$ & $R_1,R_3:\,(\pm 1,0)^4 $ \\
\hline $B_1,B_3:\,(0,\pm 1)$ & $B_1,B_3:\,(0,\pm 1)$ \\
\hline $C_1,C_2:\,(\pm 2,\pm 1)$ & $C_1,C_2:\,(\pm 2,\pm 1) $ \\
\hline $V_1,V_4:\,(\pm 2,\pm 1)$ &$V_4,V_1:\,(\mp 2,\mp 1)$ \\
\hline
\end{tabular}
\end{center}
\caption{\footnotesize Different sections of the singularities}
\label{bbbb}
\end{table}

We conclude that the low energy degrees of freedom form $SU(2)$ doublets;
 in the $SU(2)\times U(1)$ basis they are:
\begin{eqnarray}
&(R_1,R_3):\hspace{10pt}(\pm1,1;0,0)^4;\nonumber\\
&(B_1,B_3):\hspace{10pt}(0,2;\pm1,0);\nonumber\\
&(C_1,C_2):\hspace{10pt}(\pm2,0;\pm1,0);\nonumber\\
&(V_1,V_4):\hspace{10pt}(\pm2,2;\pm1,0).
\end{eqnarray}

\section{Low-Energy Coupling Constant}

\begin{figure}[h]
\begin{center}
\input{figuretta.pstex_t}
\end{center}
\caption{\footnotesize $\tilde{\alpha}$ and $\tilde{\beta}$ cycles used in the explicit calculation
of $\tau_{ij}$.}
\label{figuretta}
\end{figure}
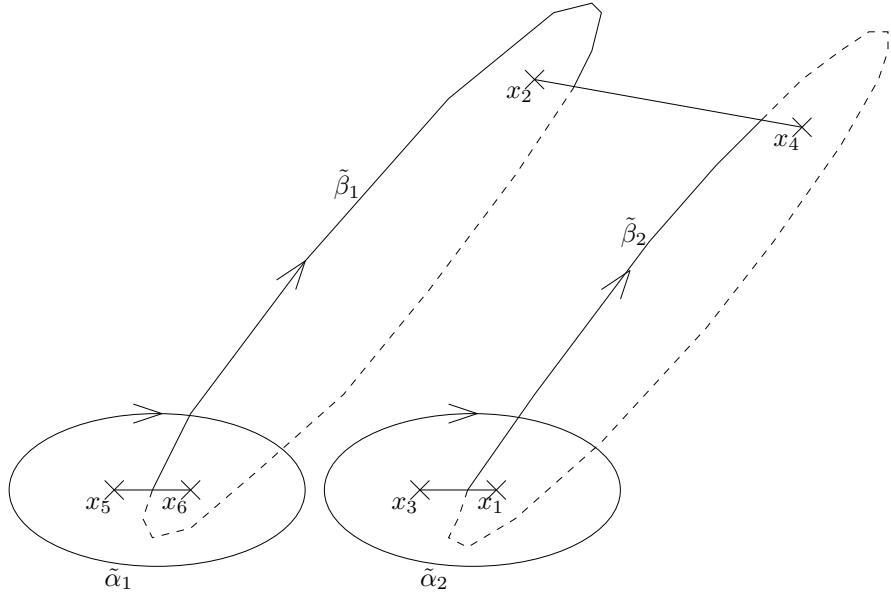

Our aim is now to determine the form of the
$U(1)^2$ low energy coupling constants near the singularity point.
They are given by the symmetric $\tau_{ij}$ matrix of the associated Riemann surface.
Setting $ U=-2+u, V = v $ the curve in the massless case assumes the following form:
\beq y^2= x(x^2-U x -V)^2-4 x^3=x(x^2-u x -v)(x^2-(u-4)x-v).
\eeq
The branch points in the $x$ plane are the following:
\beq x_3=0, \hspace{10pt} x_{2,4}=\frac{u \pm \sqrt{u^2+4 v} }{2},     \eeq
\beq x_{1,5}=\frac{(u-4) \pm \sqrt{(u-4)^2+4 v} }{2}.     \eeq
In the limit $ u, v \rightarrow 0$ the points
$x_1$, $x_2$, $x_3$, $x_4 $  collide in $x=0$;
 approximately, \beq x_1 \simeq \frac{u^2+4 v}{16}, \hspace{10pt} x_5 \simeq -4. \eeq
Introducing the new variables \beq \omega^2=u^2+4 v , \hspace{10pt} \alpha=\frac{u}{\omega}, \eeq
one has: \beq v=\omega^2(1-\alpha^2);\hspace{10pt} u=\omega \alpha.\eeq
The colliding branch points become:
\beq x_3=0, \hspace{10pt}x_{2,4}=\frac{(\alpha \pm 1) \omega }{2}, \hspace{10pt}x_1=\frac{\omega^2}{16}. \eeq
The matrix $\tau_{ij}$ is a conformally invariant quantity; so one can multiply
by $\omega$ the positions of the branch points on the $x$ plane
without altering it.
We have:
\beq x_3=0, \hspace{10pt} x_{2,4}= \frac{(\alpha \pm 1)}{2},
 \hspace{10pt} x_1= \frac{\omega}{16},\hspace{10pt} x_5=-\frac{4}{\omega}, \hspace{10pt} x_6=\infty. \eeq
In the limit $\omega \rightarrow 0$ one gets two couples of colliding branch points (at $0$ and at $\infty$).
One can now translate these points by the quantity $a$ and then apply the transformation $x \rightarrow \frac{1}{x}$.
These transformations have no effect on the conformally invariant quantity $\tau_{ij}$.
The branch points become:
\beq x_3=\frac{1}{a}, \hspace{10pt} x_1=\frac{1}{a+\omega/16},
\hspace{10pt} x_{2,4}= \frac{2}{(2 a + \alpha \pm 1)} ,
 \hspace{10pt} x_5= \frac{\omega}{-4+\omega a}, \hspace{10pt} x_6=0. \eeq
Fixing $a=1$ and keeping only the lowest order terms in $\omega \rightarrow 0$:
\beq x_3=1, \hspace{10pt} x_1=1-\frac{\omega}{16},
\hspace{10pt} x_{2,4}= \frac{2}{(2 + \alpha \pm 1)} ,
 \hspace{10pt} x_5= -\frac{\omega}{4}, \hspace{10pt} x_6=0. \eeq
The couples of colliding branch points are now at $0$ and at $1$; there are no
branch points at infinity and this makes the computation of the integrals
around the $\tilde{\alpha}$ and the $\tilde{\beta}$ cycles easier in the basis shown in figure \ref {figuretta}.

If one calculates the integral of the
holomorphic forms $\omega_1=\frac{dx}{y}$, $\omega_2=\frac{x dx}{y}$ around the $\tilde{\alpha}$
and $\tilde{\beta}$ cycles:
\beq
M_1=\left(\begin{array}{cc} \oint_{\tilde{\alpha}_1} \omega_1 &\oint_{\tilde{\alpha}_2} \omega_1\\
\oint_{\tilde{\alpha}_1} \omega_2& \oint_{\tilde{\alpha}_2} \omega_2 \end{array}\right), \hspace{20pt}
M_2=\left(\begin{array}{cc} \oint_{\tilde{\beta}_1} \omega_1 &\oint_{\tilde{\beta}_2} \omega_1\\
\oint_{\tilde{\beta}_1} \omega_2& \oint_{\tilde{\beta}_2} \omega_2 \end{array}\right);
\eeq
then $\tau_{ij}$ is given by:
\beq \tau_{ij}=M_2^{-1} M_1. \eeq
Explicit calculations give:
\beq
M_1=\left(\begin{array}{cc} - \frac{2 \pi i}{\sqrt{\frac{4}{(1+\alpha)(3+\alpha)}}} &
\frac{2 \pi i}{\sqrt{(1-\frac{1}{1+\alpha})(1-\frac{1}{3+\alpha})}}\\
0 & \frac{2 \pi i}{\sqrt{(1-\frac{1}{1+\alpha})(1-\frac{1}{3+\alpha})}} \end{array}\right),
\eeq
\beq
M_2=\left(\begin{array}{cc} - \frac{-\log(\omega)}{\sqrt{\frac{4}{(1+\alpha)(3+\alpha)}}} &
\frac{-\log(\omega)}{\sqrt{(1-\frac{1}{1+\alpha})(1-\frac{1}{3+\alpha})}}\\
0 & \frac{-\log(\omega)}{\sqrt{(1-\frac{1}{1+\alpha})(1-\frac{1}{3+\alpha})}} \end{array}\right)
\eeq
and
\beq \tau_{ij}=\left(\begin{array}{cc} \frac{2 \pi i}{-\log(\omega)} & 0 \\
0 & \frac{2 \pi i}{- \log(\omega)} \end{array}\right). \eeq
 $M_1$ is exact, calculated using residues. In $M_2$ only the most
divergent piece in the integral is kept; the $0$ in $M_2$ is a finite non-divergent piece.
So one has a non-diagonal correction to $\tau_{ij}$ of order $1/(log(\omega)^2)$, 
which is difficult to compute.

Now it is easy to write $\tau_{ij}$ in our original
basis. The matrix which transforms between the two bases is the following:
\beq
\left(\begin{array}{c}\tilde{\alpha}_1\\\tilde{\alpha}_2\\\tilde{\beta}_1\\\tilde{\beta}_2\end{array} \right)=
\left(\begin{array}{cccc}0&1&0&0\\1&-1&1&0\\-1&1&0&1\\-1&0&0&0\end{array} \right)
\left(\begin{array}{c}\alpha_1\\\alpha_2\\\beta_1\\\beta_2\end{array} \right)=S
\left(\begin{array}{c}\alpha_1\\\alpha_2\\\beta_1\\\beta_2\end{array} \right).
\eeq
If the cycles transform with the symplectic matrix
\beq M= \left(\begin{array}{cc}A&B\\C&D \end{array}\right), \eeq
$\tau_{ij}$ transforms as
\beq \tau \rightarrow (A \tau+B) (C \tau + D)^{-1}. \label{cambiotau}\eeq
In the original basis one gets (using (\ref{cambiotau}) with $M=S^{-1}$):
\beq \tau_{ij}=\left(\begin{array}{cc}-1+ \frac{4 \pi i}{-\log(\omega)} & \frac{2 \pi i}{-\log(\omega)} \\
\frac{2 \pi i}{-\log(\omega)} & \frac{2 \pi i }{- \log(\omega) } \end{array}\right), \eeq
and transforming into the  $SU(2) \times U(1)$ basis with  $M=(T^T)^{-1}$ (see $T$ in (\ref{cambio})) one finds:
\beq \tau_{ij}=\left(\begin{array}{cc}-1/2+\frac{i \pi}{-2 \log(\omega)}& \frac{\pi}{-4 \log(\omega)}  \\
\frac{\pi}{-4 \log(\omega)} & \frac{i \pi}{-2 \log(\omega)} \end{array}\right). \eeq
At the end, in the $\omega \rightarrow 0$ limit, in the $SU(2)\times U(1)$ basis it turns out that
\beq \tau_{ij}=\left(\begin{array}{cc}-1/2& 0 \\
0 & 0 \end{array}\right). \eeq

The $\beta$-function cancellation mechanism
found in \cite{AD} and in \cite{AGK} does not generalized in an obvious way to this vacuum. We found above four $SU(2)$ doublets of nonabelian
monopoles, an electric doublet and two $(2,1)$ dyonic doublets. The monopoles cancel the
contribution to the $\beta$-function of the nonabelian dual $SU(2)$ gauge bosons.
The one-loop contributions of the other degrees of freedom would cancel each other for $\tau_{11}$ such that (see \cite{AD},\cite{AGK}):
\beq  \label{condtau} \sum_i (q_i+g_i \tau_{11})^2 =2 (1+ 2 \tau_{11})^2+1=0,\eeq
{\it i. e.,}
\beq \tau_{11}=-\frac{1}{2}+\frac{\sqrt{2}}{4} i. \eeq
This condition is not satisfied.

We observe that neglecting the contribution of the electric doublet in (\ref{condtau}) the condition becomes
\beq \tau_{11}=-\frac{1}{2}, \eeq
and it would be satisfied by this vacuum. The $\beta$-function cancellation mechanism found in \cite{AD} and \cite{AGK} can be generalized to this case only if the electric doublet decouples by some unknown reason from the low energy physics.

Evidently, further investigations are needed to understand this extremely subtle issue of nontrivial SCFT with nonabelian gauge symmetry.

\section{Confinement and Chiral Symmetry Breaking Mechanism}

The superconformal limit may be approached by first breaking
chiral symmetry explicitly by introducing unequal bare quark masses.
It is easy to see \cite{CKM} that the non-local $SU(2)$ vacuum
splits into 8 local $U(1)^2$ vacua. Adding an adjoint mass term $\mu {\textrm Tr} \Phi^2$
into each of these vacua the low energy degrees of freedom (monopoles, dyons) condense.
However, in the superconformal limit all the condensates become zero (see
the discussion of the $SU(3)$ case in \cite{AGK} for more details).

The theory in the equal mass case has a $SU(N_f)$ chiral symmetry group, realized
in the fundamental representation multiplets $Q$, $\tilde{Q}^*$.
The massless case is particular: in this case the chiral symmetry
is enanced to $SO(2 N_f)$ due to the presence of generators which mix $Q$ and $\tilde{Q}^*$.

The dynamical chiral symmetry breaking has been determined in \cite{CKM} by studying
the theory at  large $\mu$ ($\gg \Lambda$). The result is that $SO(2 N_f)$ is dynamically
broken to $U(N_f)$. On the other hand, in our analysis we have found that the low
energy degrees of freedom of our theory are a flavour quartet of monopoles
and three flavour singlets dyons. It seems natural to assume that the chiral symmetry breaking is caused by the following
flavour monopole condensation:
\beq  \langle M_i \tilde{M}^i \rangle = v. \eeq
This seems to be the only way to perform the breaking $SO(8) \rightarrow U(4)$
with our low energy degrees of freedom. This is also reasonable because our monopoles are strongly coupled.

As in the $SU(3)$ case \cite{AGK}, this suggests strongly that the mechanism of confinement and chiral symmetry breaking involves strongly coupled nonabelian monopoles and dyons.

\section*{Acknowledgments}

We are grateful to Kenichi Konishi, Jarah Evslin, Stefano Bolognesi for many useful discussions.
Some of the algebraic analysis needed in this work  have been done by using   Mathematica 4.0.1  (Wolfram Research).

\appendix

\section{Monodromies}
\label{append}

Evaluating numerically the four monodromies $R_1$, $B_1$, $C_1$, $V_1$ one gets:
\begin{eqnarray}
R_1=\left(\begin{array}{cccc}1&0&0&0\\0&1&0&0\\-4&0&1&0\\0&0&0&1\end{array}\right), & B_1=\left(\begin{array}{cccc}0&-1&1&0\\0&1&0&0\\-1&-1&2&0\\-1&-1&1&1\end{array}\right)\\
C_1=\left(\begin{array}{cccc}2&-1&1&0\\0&1&0&0\\-1&1&0&0\\1&-1&1&1\end{array}\right), & V_1=\left(\begin{array}{cccc}2&0&1&0\\0&1&0&0\\-1&0&0&0\\0&0&0&1\end{array}\right).
\end{eqnarray}
The topology of the four rings yields the relations
\begin{eqnarray}
\label{check}
R_2=B_1^{-1}R_1B_1,&R_3=V_4^{-1}R_2V_4,\nonumber\\
R_4=B_3^{-1}R_3B_3,&R_1=V_1^{-1}R_4V_1,\nonumber\\
\nonumber\\
B_2=V_4^{-1}B_1V_4,&B_3=R_3^{-1}B_2R_3,\nonumber\\
B_4=V_1^{-1}B_3V_1,&B_1=R_1^{-1}B_4R_1,\nonumber\\
\nonumber\\
C_2=V_4^{-1}C_1V_4,&C_1=V_1^{-1}C_2V_1,\nonumber\\
\nonumber\\
V_2=R_1^{-1}V_1R_1,&V_3=B_1^{-1}V_2B_1,\nonumber\\
V_4=C_1^{-1}V_3C_1,&V_5=R_3^{-1}V_4R_3,\nonumber\\
V_6=B_3^{-1}V_5B_3,&V_1=C_2^{-1}V_6C_2.
\end{eqnarray}

With these relations, one gets all of the monodromies, and re-obtain the initial ones
with the last relation of each ring: this is a non trivial check of the monodromies.
The check succeeds, and the monodromies are:

\begin{eqnarray}
R_2=\left(\begin{array}{cccc}1&-4&4&0\\0&1&0&0\\0&0&1&0\\0&-4&4&1\end{array}\right), & R_3=\left(\begin{array}{cccc}1&0&0&0\\0&1&0&0\\-4&4&1&0\\4&-4&0&1\end{array}\right)\nonumber\\
R_4=\left(\begin{array}{cccc}1&0&4&0\\0&1&0&0\\0&0&1&0\\0&0&0&1\end{array}\right),&B_2=\left(\begin{array}{cccc}4&-3&1&0\\0&1&0&0\\-9&9&-2&0\\9&-9&3&1\end{array}\right),\nonumber\\
B_3=\left(\begin{array}{cccc}0&1&1&0\\0&1&0&0\\-1&1&2&0\\1&-1&-1&1\end{array}\right), & B_4=\left(\begin{array}{cccc}4&-1&1&0\\0&1&0&0\\-9&3&-2&0\\3&-1&1&1\end{array}\right),\nonumber\\
V_2=\left(\begin{array}{cccc}-2&0&1&0\\0&1&0&0\\-9&0&4&0\\0&0&0&1\end{array}\right),&V_5=\left(\begin{array}{cccc}-2&2&1&0\\0&1&0&0\\-9&6&4&0\\-6&-4&-2&1\end{array}\right),\nonumber\\
V_3=V_4=\left(\begin{array}{cccc}2&-2&1&0\\0&1&0&0\\-1&2&0&0\\2&-4&2&1\end{array}\right),&C_2=C_1,\,V_6=V_1.
\end{eqnarray}

Charges (\ref{charges}) can be calculated from these monodromies.

\end{document}

%% file: figuretta.pstex_t
\begin{picture}(0,0)%
\includegraphics{figuretta.pstex}%
\end{picture}%
\setlength{\unitlength}{3158sp}%
\begingroup\makeatletter\ifx\SetFigFont\undefined%
\gdef\SetFigFont#1#2#3#4#5{%
  \reset@font\fontsize{#1}{#2pt}%
  \fontfamily{#3}\fontseries{#4}\fontshape{#5}%
  \selectfont}%
\fi\endgroup%
\begin{picture}(6920,4645)(368,-4769)
\put(3601,-4711){\makebox(0,0)[lb]{\smash{\SetFigFont{10}{12.0}{\rmdefault}{\mddefault}{\updefault}{$\tilde{\alpha}_2$}%
}}}
\put(976,-4111){\makebox(0,0)[lb]{\smash{\SetFigFont{10}{12.0}{\rmdefault}{\mddefault}{\updefault}{$x_5$}%
}}}
\put(1576,-4111){\makebox(0,0)[lb]{\smash{\SetFigFont{10}{12.0}{\rmdefault}{\mddefault}{\updefault}{$x_6$}%
}}}
\put(3376,-4111){\makebox(0,0)[lb]{\smash{\SetFigFont{10}{12.0}{\rmdefault}{\mddefault}{\updefault}{$x_3$}%
}}}
\put(4051,-4111){\makebox(0,0)[lb]{\smash{\SetFigFont{10}{12.0}{\rmdefault}{\mddefault}{\updefault}{$x_1$}%
}}}
\put(5176,-2011){\makebox(0,0)[lb]{\smash{\SetFigFont{10}{12.0}{\rmdefault}{\mddefault}{\updefault}{$\tilde{\beta}_2$}%
}}}
\put(6376,-1261){\makebox(0,0)[lb]{\smash{\SetFigFont{10}{12.0}{\rmdefault}{\mddefault}{\updefault}{$x_4$}%
}}}
\put(4276,-886){\makebox(0,0)[lb]{\smash{\SetFigFont{10}{12.0}{\rmdefault}{\mddefault}{\updefault}{$x_2$}%
}}}
\put(1126,-4711){\makebox(0,0)[lb]{\smash{\SetFigFont{10}{12.0}{\rmdefault}{\mddefault}{\updefault}{$\tilde{\alpha}_1$}%
}}}
\put(2926,-1636){\makebox(0,0)[lb]{\smash{\SetFigFont{10}{12.0}{\rmdefault}{\mddefault}{\updefault}{$\tilde{\beta}_1$}%
}}}
\end{picture}

%% file: usp4-final.bbl
\begin{thebibliography}{100}

\bibitem{SW1}
N. Seiberg and E. Witten,   {\bf   Nucl. Phys. B426} (1994) 19; Erratum
\textit{ibid.}     {\bf   Nucl.Phys.   B430} (1994) 485, hep-th/9407087.

\bibitem{SW2}
N. Seiberg and E. Witten, {\bf Nucl. Phys.  B431} (1994) 484,
   hep-th/9408099.

\bibitem{APS}
P. C. Argyres, M. R. Plesser and N. Seiberg, Nucl. Phys. {\bf B471}
(1996)
159, hep-th/9603042;

\bibitem{CKM}
G. Carlino, K. Konishi and H. Murayama,
   {\bf  JHEP   0002}  (2000) 004,     hep-th/0001036;
 {\bf    Nucl. Phys.  B590}  (2000) 37,     hep-th/0005076.

\bibitem{CKKM}
G. Carlino, K. Konishi, Prem Kumar  and H. Murayama,
  {\bf    Nucl. Phys.  B608    }  (2001) 51, hep-th/0104064.

\bibitem{GNO}   P. Goddard, J. Nuyts and D. Olive,   {\bf Nucl. Phys.  B125}
(1977) 1.

\bibitem{BK}   S. Bolognesi and K. Konishi,  {\bf    Nucl. Phys.  B645 }  (2002) 337, hep-th/0207161.

\bibitem{monopoli}  R. Auzzi, S. Bolognesi, 
 J. Evslin, K. Konishi, H. Murayama, in preparation.

\bibitem{AD}
P. C. Argyres and M. R. Douglas,   {\bf    Nucl. Phys.  B448}  (1995) 93,   hep-th/9505062.

\bibitem{SCFT}
P. C. Argyres,  M. R. Plesser,  N. Seiberg and E. Witten, {\bf  Nucl. Phys.
{\bf 461}}   (1996) 71, hep-th/9511154.

\bibitem{Eguchi}
 T. Eguchi,  K. Hori, K. Ito and S.-K. Yang, {\bf  Nucl. Phys. {\bf B471}}
(1996)
430, hep-th/9603002.

\bibitem{AGK}  R. Auzzi, R. Grena and K. Konishi,   {\bf Nucl. Phys.    B653 } (2003) 204, hep-th/0211282.

\bibitem{curves}

P.~C.~Argyres and A.~F.~Faraggi, {\bf Phys. Rev. Lett {\bf 74}} (1995)
3931, hep-th/9411047;
A. Klemm, W. Lerche, S. Theisen and S. Yankielowicz,  {\bf  Phys. Lett.
{\bf B344} }    (1995) 169, hep-th/9411048;
 {\bf  Int. J. Mod. Phys. A11}   (1996) 1929, hep-th/9505150;
A. Hanany
and Y. Oz,   {\bf  Nucl. Phys. {\bf B452} }   (1995) 283,
hep-th/9505075;
P.  C.  Argyres, M.  R.  Plesser and A.  D.  Shapere,   {\bf  Phys.  Rev.
Lett.  {\bf 75} }      (1995) 1699, hep-th/9505100;
P. C. Argyres and A. D. Shapere,   {\bf    Nucl. Phys. {\bf B461} }    (1996)
437,
hep-th/9509175;   A. Hanany,
 {\bf   Nucl.Phys. {\bf B466}}    (1996) 85,  hep-th/9509176.

\bibitem{ds}
M.~R.~Douglas and S.~H.~Shenker,
Nucl.\ Phys.\ B {\bf 447} (1995) 271,
hep-th/9503163.



































\end{thebibliography}
